\documentclass[%
reprint,
superscriptaddress,
showpacs,
nofootinbib,
amsmath,amssymb,
prc,
floatfix, ]%
{revtex4-1}

\usepackage{color}

\usepackage{graphicx}
\usepackage{dcolumn}
\usepackage{bm}
\usepackage[dvipdfmx,bookmarks=true,colorlinks,%
            citecolor=blue,linkcolor=blue,anchorcolor=blue,filecolor=blue,urlcolor=blue,%
           ]{hyperref}

\allowdisplaybreaks

\begin{document}


\title{
Two-body dissipation effect in nuclear fusion reactions}

\author{Kai Wen}%
 \email{k.wen@surrey.ac.uk}
 \affiliation{Faculty of Engineering and Physical Sciences,
              University of Surrey, Guildford, Surrey, GU2 7XH, United Kingdom}

\author{M. C. Barton}%
 \affiliation{Faculty of Engineering and Physical Sciences,
              University of Surrey, Guildford, Surrey, GU2 7XH, United Kingdom}

\author{Arnau Rios}%
 \affiliation{Faculty of Engineering and Physical Sciences,
              University of Surrey, Guildford, Surrey, GU2 7XH, United Kingdom}

\author{P. D. Stevenson}%
 \affiliation{Faculty of Engineering and Physical Sciences,
              University of Surrey, Guildford, Surrey, GU2 7XH, United Kingdom}
\date{\today}

\begin{abstract}

Friction coefficients for the fusion reaction
$^{16}$O+$^{16}$O $\rightarrow$ $^{32}$S are extracted based on
both the time-dependent Hartree-Fock
and the time-dependent density matrix methods.
The latter goes beyond the mean-field approximation by taking
into account the effect of two-body correlations,
but in practical simulations of fusion reactions we find that
the total energy is not conserved.
We analyze this problem and propose a
solution that allows for a clear quantification of
dissipative effects in the dynamics.
Compared to mean-field simulations,
friction coefficients in the density-matrix approach
are enhanced by about $20 \, \%$.
An energy-dependence of the dissipative mechanism is also demonstrated,
indicating that two-body collisions are more efficient
at generating friction at low incident energies.
\end{abstract}


\maketitle

\section{Introduction}

For a wide range of incident energies, the collisions
of two nuclei exhibit typical dissipative properties \cite{Bri85,SH85}.
The kinetic energy initially residing in collective
motion irreversibly converts into intrinsic nuclear excitations.
This is in analogy to the Brownian motion, where the kinetic
energy of the Brownian particle converts into
the surrounding heat bath \cite{Blocki1978,F87}.
Exploiting this analogy,
the language of non-equilibrium statistical physics
has been borrowed to describe the dissipation occurring in nuclear reactions
since the 1980s \cite{Bjornholm1982_NPA391-471,F87}.
The concept of friction has been introduced and widely accepted
in the study of heavy ion collisions \cite{Blocki1978,SH85,F87}.
Various theoretical models have been developed
to understand the mechanisms underlying this special dissipative process \cite{Abe96,Lac04,Lacroix2002,SCKSW12,Bjornholm1982_NPA391-471,Wen2013_PRC111-012501,Wen2014_PRC90-054613}.

Among these theoretical models,
the time-dependent Hartree-Fock (TDHF) approach
stands out as a general theoretical framework that allows
for a self-consistent quantal modelling of large amplitude
nuclear collective motion
\cite{BKN76,Guo2008_PRC77-041301R,Guo2007_PRC76-014601,Simenel2008,Simenel2012,Bulgac2013,NMMY16}.
TDHF has been extensively used in the past to study low-energy nuclear collisions
\cite{Cus85,Umar2006,Keser2012,Simenel2013,Simenel2013b,Simenel2004}.
Since this method relies on a mean-field or density-functional description of
nuclear dynamics, the dissipation encoded in the dynamics is
due to microscopic one-body processes \cite{Blocki1978,Randrup1980,Randrup1984}.
By mapping the TDHF evolution to the one-dimensional Langevin equation,
a method called Dissipative Dynamics TDHF (DD-TDHF)
has also been developed to study the dissipation
in nuclear fusion reactions \cite{WLA09,WL08,AWL09}.
Dissipative effects extracted from DD-TDHF
are of a one-body type,
in agreement with the idea that
dissipation is caused by the exchange of nucleons
across the window between the colliding nuclei \cite{F87}.

The exact dynamics of a quantum many-body system is governed
by equations that in principle go beyond the mean-field approach.
When projected into time-local many-body density matrices,
the dynamics can be expressed in terms of the so-called
Bogoliubov--Born--Green--Kirkwood--Yvon (BBGKY) hierarchy
\cite{Born1946,*Bogoliubov1946,*Kirkwood1946,Ring1980-book}.
Different levels of truncation within the hierarchy provide
different descriptions of the many-body dynamics and
higher order truncation schemes are expected to describe more
accurately the time evolution of the strongly correlated systems
\cite{Gong1990,Akbari2012,Akbari2012b,Tohyama2015}.
The TDHF approach arises naturally as the the lowest order
truncation scheme in the BBGKY hierarchy,
assuming that two- and higher-body correlations are negligible
\cite{WC85,Schmitt1990}.

To go beyond the mean-field approximation,
in nuclear physics another truncation scheme has been
implemented to account for dynamical effects on the
two-body density matrix
\cite{WC85,Gong1990}.
This so-called Time-Dependent Density Matrix (TDDM) approach
extends the TDHF method by including
terms that account for the evolution of the two-body density matrix
in the BBGKY hierarchy and by neglecting
three-body and higher order correlations
\cite{WC85,Tohyama1987,TM94,TU02, TU16,Tohyama2015}.
The numerical cost associated to directly solving the
corresponding set of TDDM coupled equations is large
even with the presently available computational power.

A practical method has been suggested and applied to reduce
this numerical task. One can expand the problem into
a single-particle basis that
evolves following a TDHF-like equation \cite{WT78,K80,AG10}.
All the one- and two-body observables
are then built up using this moving basis set,
truncated at a given maximum number of states.
This numerical technique facilitates the calculation significantly,
and provides a conserving approximation in the sense that it formally
conserves particle number as well as total energy over time \cite{WC85,Gong1990,Akbari2012}.
It has been successfully implemented
in the study of nuclear ground-state properties
in a self-consistent three-dimensional setting
\cite{Barton2018,Barton2018b}.
However, in practical simulations of
large amplitude collective motion,
we find that this technique comes at the price of losing energy conservation~\cite{Gong1990},
due to the incompleteness of basis.
This drawback hinders a quantitative study on
the dissipation mechanism in the framework of TDDM.

In this work, we present a strategy that simultaneously
probes the problem of basis incompleteness and restores
the conservation of energy.
This strategy is implemented in a practical setting that
can easily be extended to other methods
and that allows the extraction of information
on dissipation in the system even if the total energy
is only partially conserved.
With this method, we quantify the effect of two-body
dissipation in a symmetric fusion reaction of two oxygen isotopes.

The paper is organized as follows.
In Sec.~\ref{sec:theo},
we give the formulation of the basic TDDM equations,
introduce the numerical method to restore
the conservation of total energy and discuss a macroscopic reduction
procedure to extract friction coefficients.
We apply these methods to study dissipative process in the fusion reaction
$^{16}$O+$^{16}$O $\leftrightarrow$ $^{32}$S
in Sec.~\ref{sec:path}.
A summary and concluding remarks are given in Sec.~\ref{sec:sum}.

\section{\label{sec:theo}Theoretical framework}

In the application of TDDM,
a full calculation of the two-body-interaction matrix requires
a large numerical effort at every time step \cite{Barton2018}.
This numerical cost also increases rapidly
with the number of basis states and precludes realistic
applications for intermediate mass nuclei.
In consequence, we have adopted the TDDM$^{\rm P}$ approximation,
where only the interaction between time-reversed pairs is considered.
This simplified implementation of TDDM provides a generalization
of pairing dynamics and has been successfully applied
to study the effect of nuclear correlations
on the breakup mechanism of light nuclides \cite{Assie2008,AL09}.
We adopt the TDDM$^{\rm P}$ approximation in our calculation and devote
the next subsection to recapitulate the basic formulation of the
TDDM and TDDM$^{P}$ methods.
Further details can be found in references~\cite{TM94,AL09}.
\subsection{\label{subsec:theoA}TDDM$^{P}$ implementation of TDDM}

The TDDM method aims at determining the time evolution of
both the one body density matrix, $\rho$, and two-body correlation
matrix, $C_{2}$, in a self-consistent way,
assuming three-body and higher-order correlations are negligible.
$C_{2}$ is customarily defined as the correlated
part of the two-body density matrix,
$C_{2} = \rho_{2}-\hat A(\rho\rho)$,
where $\hat A$ stands for an antisymmetrization operator.

To solve the first two equations of the BBGKY hierarchy,
we chose to expand $\rho$ and $C_2$ in a finite
number of single particle states,
${\psi_{i}({\bf r},t), \, i=1,\cdots,N_\text{max}}$,
which evolve in time obeying a TDHF-like equation of motion,
\begin{align}
i\hbar  \dot \psi_{i}({\bf r},t)=
    \hat{h}(t,\rho )\psi_{i}({\bf r},t) \, .\label{eq:tddm1}
\end{align}
We note that the mean-field Hamiltonian, $\hat{h}(t,\rho)$, depends on the
correlated one-body density matrix.
The TDHF evolution would, instead, rely on an uncorrelated Hartree-Fock-like density.
Note also that the truncation parameter $N_\text{max}$ is introduced here.

In terms of this moving basis, the one-body density and
two-body correlation matrices are expressed as
\begin{widetext}
\begin{align}
\rho({\bf r_1},{\bf r_{1'}}; t) &=\sum_{ii'} n_{ii'}(t)\psi_{i}({\bf r_1},t)\psi_{i'}^{*}({\bf r_{1'}},t),\\
C_{2}({\bf r_1},{\bf r_2},{\bf r_{1'}},{\bf r_{2'}};t) &=\sum_{iji'j'} C_{iji'j'}(t)
                 \psi_{i}({\bf r_{1}},t)\psi_{j}({\bf r_{2}},t)\psi_{i'}^{*}({\bf r_{1'}},t)\psi_{j'}^{*}({\bf r_{2'}},t) ,\,
\end{align}
\end{widetext}
where all indices run over the whole basis set.
In addition to Eq.~(\ref{eq:tddm1}),
the TDDM methods solve the dynamics in terms
of the time evolution of single-particle occupation numbers,
$n_{ii'}(t)$, and the correlation matrix, $C_{iji'j'}(t)$,
\begin{align}
i\hbar\dot{n}_{ii'}(t) &= \sum_{jkl}\left[ \langle ij|v|kl\rangle C_{kli'j}
                       -C_{ijkl}(t) \langle kl|v|i'j\rangle\right],  \label{eq:tddm2} \\
i\hbar\dot{C}_{iji'j'}(t) & =B_{iji'j'}(t)+P_{iji'j'}(t)+H_{iji'j'}(t), \label{eq:tddm3}
\end{align}
where $\langle ij|v|kl \rangle$ is a two-body interaction matrix element.
The matrix terms on the right hand-side of
Eq.~(\ref{eq:tddm3}) represent different correlation mechanisms.
$B_{iji'j'}$ is generally associated with the Born terms,
containing the physics of direct in-medium collisions \cite{Gong1990}.
The terms $P_{iji'j'}$ and $H_{iji'j'}$ 
represent higher-order correlations.
All these terms can be expressed as a combination
of interaction matrix elements,
occupation numbers and correlation matrices \cite{Gong1990},
forming a closed set of TDDM equations.

There are two major computational bottlenecks in
the practical implementation of the TDDM equations.
One is calculation of the interaction matrix elements
in Eq. (\ref{eq:tddm2}), which requires in principle a loop
over 4 different single-particle indices at every time step.
The second bottleneck arises similarly in the
solution of Eq.~(\ref{eq:tddm3}),
which requires manipulations of a 4-index tensor
of size $N_\text{max}^4$ \cite{Barton2018}.
A significant reduction of the numerical cost can be achieved
by the TDDM$^{P}$ implementation of the TDDM equations
\cite{Assie2008,AL09}.
In this approach, one assumes that the residual
two-body interaction is dominated by time-reversed
pair states, $\{i, \bar i \}$.
One keeps only the elements of the interaction
matrix between these pairs,
$\langle i \bar i |v| j \bar j \rangle$,
and assumes that all other matrix elements are zero.
The correlation matrix $C$ is also only
formed by time-reversed pair states.
The number of required matrix elements in both
$V$ and $C$ is therefore significantly reduced.
Further, the term $H_{iji'j'}$ in Eq. (\ref{eq:tddm3}) cancels out,
and Eqs~(\ref{eq:tddm2}) and (\ref{eq:tddm3}) reduce to
\begin{align}
\dot{n}_{\alpha}&=\frac{2}{\hbar}\sum_{\gamma} \text{Im}( V_{\alpha\gamma}C_{\gamma\alpha}),\label{eq:tddmp2}\\
i\hbar\dot{C}_{\alpha\beta}&= V_{\alpha\beta}\left[ (1-n_{\alpha})^{2}n^{2}_{\beta}
                -(1-n_{\beta})^{2}n^{2}_{\alpha} \right] \label{eq:tddmp3} \\
                &+\sum_{\gamma}V_{\alpha\gamma}(1-2n_{\alpha})C_{\gamma\beta}
                -\sum_{\gamma}V_{\gamma\beta}(1-2n_{\beta})C_{\alpha\gamma}.
                \nonumber
\end{align}
Here, the Greek indexes $\alpha,\beta,\ldots$ represent
a pair of time reversed states.
$V_{\alpha\beta}$ is the antisymmetric element of
the interaction matrix between two pairs,
$V_{\alpha\beta}=\langle \alpha\bar{\alpha}|v|\beta\bar{\beta}\rangle_{A}$,
and $C_{\alpha\beta}$ is the corresponding two-body correlation tensor,
$C_{\alpha\beta}=\langle \alpha\bar{\alpha}|C|\beta\bar{\beta}\rangle$.
In a sense, the interaction adopted here can be seen
as a generalized BCS interaction, and TDDM$^P$ is
akin to a superfluid time-dependent approach \cite{Assie2008}.


\subsection{\label{subsec:theob}TDDM with optimized basis}

The conservation of total energy is critical for
the analysis of dissipation mechanism.
Without the conservation of total energy,
the collective kinetic and potential energies
can not be assigned unambiguously and
the quantification of dissipation processes
becomes impossible.

\begin{figure}
\begin{centering}
\includegraphics[width=0.90\columnwidth]{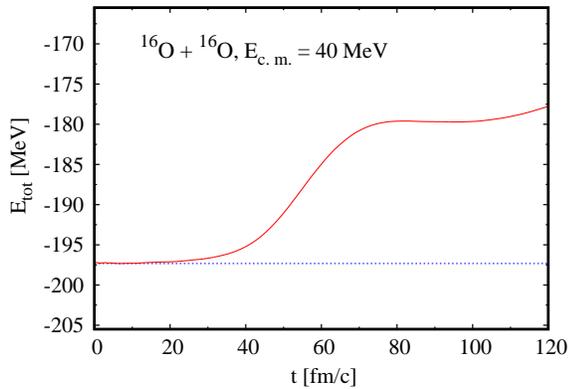}
\par\end{centering}
\caption{\label{fig:nc}(Color online)
Total energy of the fusion system $^{16}$O+$^{16}$O $\rightarrow$ $^{32}$S
at incident $E_{\rm c. m.} = 40$ MeV as a function of time $t$.
The calculation is performed using TDDM$^{\rm P}$ model.
The dashed blue line indicates the constant initial energy for reference.
This figure is obtained with $N_\text{max}=60$.
}
\end{figure}

Formally, both the TDDM and TDDM$^P$
equations preserve the conservation of average particle number, momentum and
energy if a complete basis is present~\cite{Gong1990,Assie2008}.
However, a finite basis evolving with
Eq.~(\ref{eq:tddm1}) spoils the energy conservation in practical calculations~\cite{Gong1990}.
The violation of energy conservation depends sensitively
on the form and strength of the residual interaction, and can not be
remedied by increasing the size of the model space, $N_\text{max}$,
within practical limits.
%
Fig.~\ref{fig:nc} shows an example of a TDDM$^{\rm P}$
simulation of the collision $^{16}$O+$^{16}$O $\rightarrow$ $^{32}$S
at an incident energy of $E_{\rm c. m.} = 40$ MeV.
In the approaching phase, $t < 30$ fm/c,
the energy is approximately conserved.
A rapid increase in energy is observed in the region $30$ fm/c $< t < 80$ fm/c,
which corresponds to the merging process:
starting from the point of contact to the formation of a
compact system. In this example, the total energy increases by about $18$ MeV.
This corresponds to a relative deviation compared to the initial
value of about $10 \, \%$.
We take this as an indication that
the moving basis that reproduces well
the projectile and target nuclides in the initial state
is not reliable in the rapidly evolving fusion process.

The inclusion of a complete set of basis states
to simulate nuclear fusion reactions is, however, infeasible.
The idea to adopt a set of moving basis whose
time evolution obey TDHF-like equations is based on the
expectation that this basis
would to some extent satisfy the requirement
of the actual solution of the full dynamics.
Keeping the same motivation,
we modify Eq.~(\ref{eq:tddm1})
by introducing an additional term in the mean-field Hamiltonian,
\begin{align}
i\hbar \dot \psi_{i}({\bf r},t)
         =\left[\hat{h}(t,\rho)+v'_{i}(t)\right]\psi_{i}({\bf r},t) \, .\label{eq:tddm0}
\end{align}
The gradient of the correction terms $v'_{i}(t)$ is related to the
average momentum of each single-particle state,
\begin{eqnarray}
\nabla v'_{i}(t) = \beta(t) \langle \psi_{i}(t)|\hat{p}|\psi_{i}(t) \rangle, \label{betain}
\end{eqnarray}
where $\hat{p}$ is the momentum operator.
$\beta(t)$ is a free parameter that allows one to fix the scale of the correction.
An overall constant in $v'_i$ will not change the result, and we choose to set it to zero.

The terms $v'_{i}(t)$ are designed
to optimize the basis by conserving the
total energy upon adjusting the parameter $\beta(t)$ at each time step.
Whenever $\beta=0$, no adjustment is necessary and the energy is
conserved.
A non-zero $\beta$ will appear when the basis of Eq.~(\ref{eq:tddm1})
fails to conserve the total energy, with
a larger $\beta$ in principle indicating a worse-performing basis.
Thus, $\beta(t)$ can also be seen as
a proxy that quantifies to what extent the
moving basis defined by Eq.~(\ref{eq:tddm1}) is satisfactory,
in the sense that it provides energy conservation.
We note that while $v'_i$ depends on the orbit $i$, $\beta$ is
assumed to be the same for all orbits. Because of this orbital dependence,
$v'_{i}$ cannot be absorbed in a redefinition of the mean-field hamiltonian.

To fix $\beta(t)$, the following numerical procedure
is performed.
At an arbitrary time, $t_0$, we evolve the system for one time step,
$\Delta t$, in two 
independent ways. The first follows the TDHF-like trajectory of Eq.~(\ref{eq:tddm1}).
The second follows Eq.~(\ref{eq:tddm0}), with a
small $\beta(t_{0})= \beta'$, which is arbitrarily set to $10^{-4}$ c/fm
in this work.
If energy is not conserved, after a time step
$\Delta t$ the energies of the two trajectories can be different.
The total energy of the first trajectory
changes  from $E(t_{0})$ to $E_{1}(t_{0}+\Delta t)$,
whereas
the total energy of the second trajectory
changes to $E_{2}(t_{0}+\Delta t)$.
When $\Delta t$, $\beta'$,
as well as the finally desired $\beta(t_{0})$ are all small,
$E_{2}(t_{0}+\Delta t)-E_{1}(t_{0}+\Delta t)$
is proportional to $\beta'$ by a constant.
$\beta(t_{0})$ can be fixed using this linear relation as
\begin{eqnarray}
\beta(t_{0}) = \beta' \frac{E_{1}(t_{0}+\Delta t)-E(t_{0})}
                  {E_{2}(t_{0}+\Delta t)-E_{1}(t_{0}+\Delta t)}.
\end{eqnarray}
Having obtained $\beta(t_{0})$, 
we restart the time evolution from
time $t_{0}$ following Eq. (\ref{eq:tddm0}).
With the choice of $\beta(t_0)$ above,
the total energy will be conserved up to $t_{0}+\Delta t$,
$E(t_{0}+\Delta t)=E(t_{0})$.
Repeating this procedure at $t_{0}+\Delta t$, $t_{0}+2\Delta t ...$, we find
a constant total energy
$E(t_{0})=E(t_{0}+\Delta t)=E(t_{0}+2\Delta t)...$,
while repeatedly adjusting $\beta(t)$ as a function of time.
The cost of performing this procedure is obviously about a factor of two
heavier than the original solution.

This strategy provides a practical implementation of
energy-conserving TDDM equations.
It also allows for a clear quantification of energy non-conserving dynamics
whenever $\beta \neq 0$.
In principle, the correction introduced
in Eq.~(\ref{eq:tddm0}) should also change the form
of Eqs.~(\ref{eq:tddm2}) and (\ref{eq:tddm3}).
However, our simulations indicate that $\beta(t)$ is rather small, so
we keep the form of these equations unchanged.

\subsection{\label{subsec:theoc}Macroscopic reduction procedure}

For simplicity, we consider a head-on symmetric collision along the $z$ axis.
In the center-of-mass coordinate frame,
we keep the identities of both the projectile and target. In other words,
the projectile and target can be identified by summing over
the single-particle states that were originally ascribed to each one of them.
The collective coordinate $R$ at time $t$ is
defined as the relative distance between
the center-of-masses of projectile and target,
\begin{align}
R(t) &= \langle\Psi(t)_{\rm pro}|z|\Psi(t)_{\rm pro}\rangle
      -\langle\Psi(t)_{\rm tar}|z|\Psi(t)_{\rm tar}\rangle.
\end{align}
In the case of TDHF, $\Psi(t)$ represents a single Slater determinant formed
of all the occupied single-particle states, $\psi_i$.
For TDDM, 
the center-of-masses of the projectile or target can be
expressed as
\begin{align}
\langle & \Psi(t)_{\rm pro(tar)}| z|\Psi(t)_{\rm pro(tar)}\rangle \nonumber\\
&=\frac{1}{N_{\rm pro(tar)}}
    \sum_{i,j \in {\rm pro(tar)}} n_{i,j} \langle\psi_{i}(t)|z|\psi_{j}(t)\rangle, \label{eq:tddmR}
\end{align}
where $N_{\rm pro(tar)}$ is the particle number of the projectile (target),
and $i$ and $j$ are indices belonging to the projectile or the target.
The expectation values of other one-body operators,
like the total momentum $P_{\rm pro (tar)}$ mentioned later,
are calculated in the same way.

As long as a one-to-one correspondence
between $R$ and $t$ exists in the fusion process,
we can label the state $\Psi$ as well as the
collective variables as a function of $R$
instead of $t$. For instance,
the variables of collective momenta,
collective kinetic energy,
and collective potential energy
can all be expressed as a function of $R$:
\begin{align}
P(R)&=P_{\rm pro} - P_{\rm tar},
\label{eq:Pdef}
        \\
T_{\rm coll}(R) &= \frac{P^{2}_{\rm pro}(R)}{2M_{\rm pro}}+\frac{P^{2}_{\rm tar}(R)}{2M_{\rm tar}},\\
V_{\rm coll}(R) &= E_{\rm tot}(R)-E_{\rm pro}(R)-E_{\rm tar}(R), \label{eq:Vdef}
\end{align}
where $P_{\rm pro}$ and $P_{\rm tar}$ are the total momentum
of projectile and target calculated in the same way as Eq. (\ref{eq:tddmR});
$M_{\rm pro}$ and $M_{\rm tar}$ are the
total mass of projectile and target;
$E_{\rm tot}$ is the total energy of the whole system and
$E_{\rm pro}$ and $E_{\rm tar}$ are the total energies of
the projectile and target.

\begin{figure}
\begin{centering}
\includegraphics[width=0.90\columnwidth]{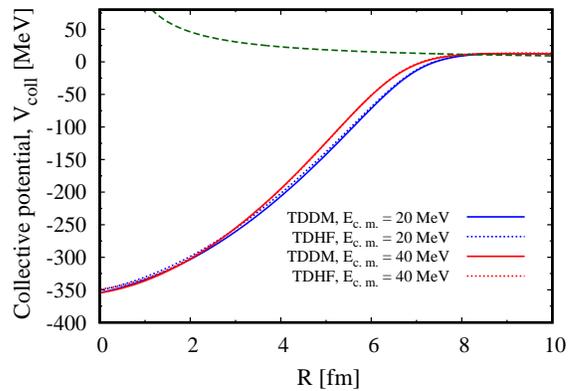}
\par\end{centering}
\caption{\label{fig:pot}(Color online)
Collective potential defined in Eq. (\ref{eq:Vdef}) as
a function of relative distance, $R$,
for the fusion path in the reaction
$^{16}$O+$^{16}$O $\rightarrow$ $^{32}$S.
The solid and dotted lines
indicate the results of TDDM and TDHF, respectively.
The blue (lower) and red (upper) lines indicate the results at incident energies of
$E_{\rm c. m.} = 20$ MeV and $E_{\rm c. m.} = 40$ MeV, respectively.
The green (dashed) line
shows the asymptotic Coulomb potential, $64e^{2}/R$, for reference.
}
\end{figure}

Figure~\ref{fig:pot} shows the collective potential
energy defined in Eq.~(\ref{eq:Vdef}) along the reaction path of the fusion reaction
$^{16}$O+$^{16}$O $\rightarrow$ $^{32}$S as a
function of relative distance $R$. At large distances,
the collective potential
agrees with the asymptotic Coulomb potential, $V_{\rm coll} \approx Z^2e^{2}/R$,
shown with a dashed (green) line for reference.
Overcoming the Coulomb barrier at $R = 8.3$ fm,
the collective potential monotonically decreases as
the nuclei come closer together and fuse.
The red and blue curves indicate two different incident energies of
$E_{\rm c. m} = 20$ MeV and $E_{\rm c. m} = 40$ MeV, respectively.
The differences in collective potentials at the two incident energies is at most of
$20$ MeV in the region $2.5$ fm $< R < 7.5$ fm.
We find that the results
at a higher incident energy are higher 
than those at lower incident energy.
As studied in references~\cite{WLA09,WL08},
this is due to the different rate of rearrangement
among the intrinsic degrees of freedom between
the fast and slow collision.
The differences between the TDHF and TDDM approaches are
much smaller than those associated with the incident energy.
The THDF collective potential is within $5$ MeV
of the TDDM potential for all positions and energies.
This indicates a relatively small effect of two-body dissipation on the
collective potentials in this reaction.

The intrinsic energy
\begin{align}
E_{\rm intr}(R) &= E_{\rm c.m.}-E_{\rm coll}(R) \, ,
\label{eq:intr}
\end{align}
is
obtained by subtracting the collective energy,
$E_{\rm coll}(R) = T_{\rm coll}(R)+V_{\rm coll}(R)$, from the
the initial bombarding
energy in the center-of-mass coordinate frame.
To define the friction force $F_{\rm fric}$,
we assume that all the work done by this force
is converted into intrinsic energy.
Under this assumption, the friction force can be extracted as the
derivative with respect to the $R$ collective variable
of the intrinsic energy,
\begin{align}
F_{\rm fric}(R) &= \frac{dE_{\rm intr}(R)}{dR} \, .
\end{align}
According to the Rayleigh formula \cite{AWL09,Frobrich1998_PR292-131},
the friction coefficient $\gamma(R)$ as a function of $R$ is
extracted from the ratio:
\begin{align}
\gamma(R) &= \frac{F_{\rm fric}(R)}{P(R)}, \label{fric}
\end{align}
with $P(R)$ defined as in Eq.~(\ref{eq:Pdef}).

\section{\label{sec:path}Application}

In this section we apply the macroscopic reduction procedure
to investigate the dissipation mechanism of the reaction
$^{16}$O+$^{16}$O $\rightarrow$ $^{32}$S.
The method to conserve the total energy
within the TDDM$^{P}$
scheme introduced above
is incorporated in the calculation. 
The TDDM$^P$ dynamics is built on top of the
Sky3D code \cite{Maruhn2014}, which solves the TDHF equations on
a three dimensional Cartesian mesh with Skyrme forces \cite{Barton2018,Barton2018b}.
To solve Eqs. (\ref{eq:tddmp2}), (\ref{eq:tddmp3}) and (\ref{eq:tddm0}),
a fourth order Runge-Kutta time propagation algorithm is used
to improve the accuracy of the solution and
the convergence of the initial states.

We adopt the Skyrme III force to calculate mean-field component of the interaction
matrix as well as the mean-field hamiltonian in Eq. (\ref{eq:tddm0}) \cite{SIII}.
Skyrme III is a
standard parameterization of the Skyme force,
in which the density dependent term as well
as the spin-orbit term are present.
The residual interaction is assumed to be of
zero-range with a linear density dependence following
the standard choice in the literature \cite{Chasman76,TU02,YMM01,DBH01},
\begin{eqnarray}
v_{12}(\vec{r}_{1},\vec{r}_{2})=v_{0} \left[1-\frac{\rho(r)}{\rho_{0}} \right]
                                \delta(\vec{r}_{1}-\vec{r}_{2}),
                                \label{eq:resi}
\end{eqnarray}
where $\rho(r)$ is the nuclear density,
and $\rho=0.16$ fm$^{-3}$ is the saturation density.
The strength of the residual interaction
$v_{0}$ is set to be $-1200 $ MeV fm$^{3}$, following
reference \cite{Chasman76}.
We discretize the mesh 
in a cubic box
of size $16.5\times16.5\times16.5$ fm$^{3}$ for the preparation
of the projectile and target,
and a rectangular box of size $16.5\times16.5\times33.0$ fm$^{3}$ for the reaction.
The mesh spacing is set to $\Delta x=1.1$  fm in all directions. The time step
is $\Delta t=0.3$ fm/c. All simulations are run for a total time of $150$ fm/c.

\subsection{\label{sec:theob} Correlated ground state of $^{16}$O}

The initial correlated ground states of the projectile and target
are generated by means of the adiabatic
switching technique \cite{Rios2011}.
A static Hartree-Fock (HF) calculation is performed
first to obtain an initial mean-field ground state.
Starting from this HF state, we switch on the residual
interaction of Eq.~(\ref{eq:resi}) adiabatically. The time-dependent
residual interaction is given by the expression
\begin{eqnarray}
v_{12}(\vec{r}_{1},\vec{r}_{2},t)= \left( 1-e^{-\frac{t^{2}}{\tau^{2}}} \right)
                        v_{12}(\vec{r}_{1},\vec{r}_{2}), \label{eq:resi2}
\end{eqnarray}
which satisfies $v_{12}(\vec{r}_{1},\vec{r}_{2},t=0)=0$ and
$v_{12}(\vec{r}_{1},\vec{r}_{2},t \to \infty)=v_{12}$. While the
residual interaction is switched on, we evolve the system following the TDDM$^{P}$
equations, Eqs.~(\ref{eq:tddm1}), (\ref{eq:tddmp2}) and (\ref{eq:tddmp3}).
In order to obtain a stationary correlated state,
the adiabatic theorem requires the interaction to be switched on slowly enough.
For $^{16}$O, we find that setting $\tau=300$ fm/c provides a good compromise.

To obtain the correlated ground state of $^{16}$O,
different scheme for model spaces have
been proposed in the TDDM model~\cite{TU02}.
In this work, we use a model space consisting of $N_\text{max}=30$ orbits,
with $N_\text{max}^{\rm n}=16$ neutron states and $N_\text{max}^{\rm p}=14$ proton
states,
so that all these single-particle states are kept bound
and evaporation is avoided during the fusion process.

\begin{figure}
\begin{centering}
\includegraphics[width=0.90\columnwidth]{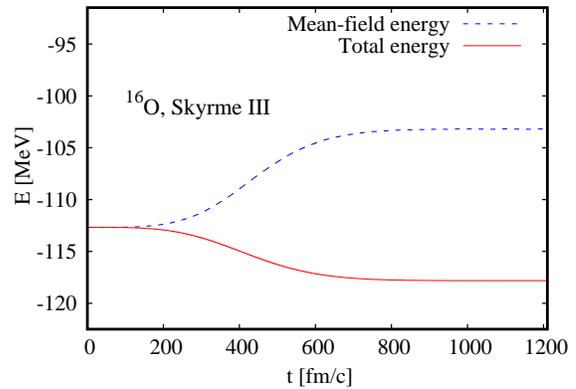}
\par\end{centering}
\caption{\label{fig:static}(Color online)
Solid line: the total energy of $^{16}$O as it evolves from the Hartree-Fock
state to the TDDM correlated state by switching on the residual
interaction adiabatically.
Dashed line: the mean-field energy in the same conditions. This figure
is obtained with $N_\text{max}=30$.
}
\end{figure}
\begin{table}[h]
	\centering
	\begin{tabular}{c|ccccc}\hline\hline
		         &  $1s$ (2)&$1p_{3/2}(4)$&$1p_{1/2}(2)$ &$1d_{5/2}(6)$&$2s$(2) \\\hline
             Neutron&    0.988 &  0.954      &  0.887       &0.068        &0.013\\\hline
             Proton &     0.992&   0.962     &   0.907      & 0.059       & \\\hline\hline
	\end{tabular}
\caption{Converged occupation numbers, $n_{ii}$,
of the single particle states for $^{16}$O.
The calculation is performed with $N_{\rm max}=30$.
The numbers in the parentheses denote the degeneracy of the corresponding orbits.}
	\label{tab:occ}
\end{table}
Figure~\ref{fig:static} shows the evolution of the total energy (solid line)
with time as the residual interaction is switched on following
Eq.~(\ref{eq:resi2}).
The dashed (blue) curve indicates the energy without
two-body correlations. The initial system is uncorrelated
and the total energy is entirely due to the mean-field contribution.
As the residual interaction is switched on, the system becomes
more bound by about $5.2$ MeV. The mean-field contribution, in contrast,
is about $9.5$ MeV less attractive, mostly due to the increase in kinetic energy
associated with correlations.

Both curves display a quite satisfactory convergence,
indicating that, as we turn on correlations adiabatically,
a stable correlated ground state is obtained.
With the parameter set of Skyrme III for the mean-field
part and Eq.~(\ref{eq:resi}) for the residual interaction,
the final contribution of the two-body correlations to the total
energy is about $14.7$ MeV for the ground state of $^{16}$O.
We note that the total energy is conserved after convergence.
In other words, $\beta(t)$ in Eq. (\ref{betain}) turns out to be zero
if the moving basis of Eq.~(\ref{eq:tddm0}) is used.
This is no longer true when the two nuclei collide
as shown in the next subsection.

Table \ref{tab:occ} gives the occupation numbers, $n_{ii}$, of different single
particle orbits for the correlated ground state of $^{16}$O.
The deeply bound $1s_{1/2}$ and $1p_{3/2}$ neutron and proton states keep more than
$95 \%$ of the single-particle occupation.
Correlations have the largest effect near the Fermi surface, where they effectively deplete the
$1p_{1/2}$ states by about $10 \%$ and allow for a $6-7 \%$ population of the
$1d_{5/2}$ states.
Neutron $2s_{1/2}$ orbits remain almost unpopulated.
We note that in the TDDM method 
the strength of two-body correlation effects and single-particle occupations depend
on the strength of the residual interaction.
A systematic study on the correlated static state will be addressed
in a separate work.

\subsection{\label{sec:theob} Dissipation with two-body correlations}

In this subsection, we explore the dissipation dynamics of the
reaction $^{16}$O+$^{16}$O $\rightarrow$ $^{32}$S.
The initial relative distance between the two
correlated ground states of $^{16}$O is set to be $10$ fm.
In the center-of-mass frame, we boost the projectile
and target symmetrically by assigning an
initial velocity to all the single-particle states.
For the TDDM$^{P}$ calculation,
the numerical procedure introduced in
Sec. \ref{sec:theob} is applied to the time evolution of the orbits.
The occupation number and the correlation matrix evolve following
Eqs.~(\ref{eq:tddmp2}) and (\ref{eq:tddmp3}).

\begin{figure}
\begin{centering}
\includegraphics[width=0.90\columnwidth]{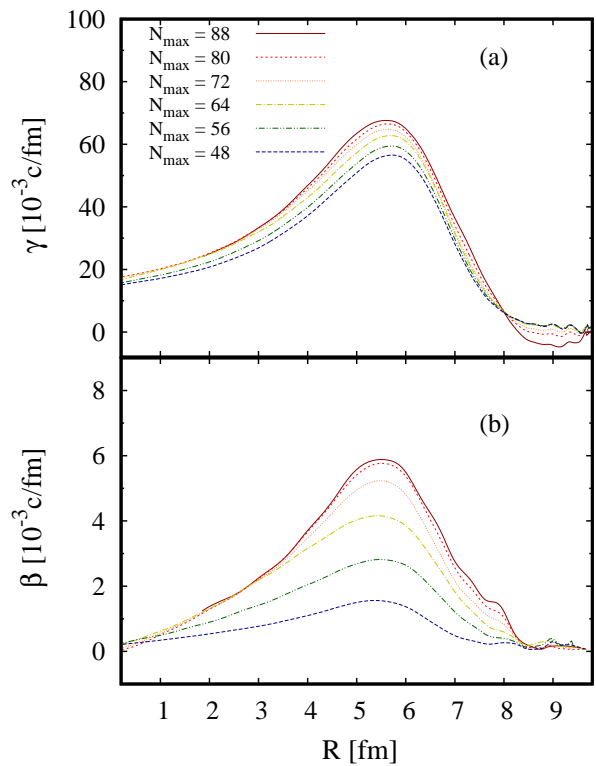}
\par\end{centering}
\caption{\label{fig:conv}(Color online)
Panel (a): friction coefficient $\gamma$ as a function of $R$ for the fusion reaction
$^{16}$O+$^{16}$O $\rightarrow$ $^{32}$S at
$E_{\rm c. m.} = 40$ MeV. Different line styles indicate results calculated
with different $N_\text{max}$.
Panel (b): $\beta$ parameters in the same conditions.
}
\end{figure}

Figure~\ref{fig:conv}(a) shows the friction coefficient $\gamma$,
defined in Eq.~(\ref{fric}), as a function of the relative distance $R$ for a
collision energy of $E_{\rm c. m.} = 40$ MeV.
The parameter $\beta$ (bottom panel) of Eq. (\ref{betain}) is shown in
Figure~\ref{fig:conv}(b).
%
Since the one-to-one correspondence between $R$ and $t$ is valid
during the collision,
$\beta$ is also expressed as a function of $R$ instead of $t$.
Curves of different styles indicate results
calculated with different total number of orbits, $N_\text{max}$,
for the total system. In other words, $N_\text{max}$ here is the sum of
$N_\text{max}$ of both projectile and target.
The initial ground states of $^{16}$O are thus constructed using $N_\text{max}/2$
orbits with $N^{\rm n}_\text{max}=N^{\rm p}_\text{max}=N_\text{max}/4$.
As $N_\text{max}$ increases,
we find that both $\beta$ and $\gamma$ converge in the region $R < 8$ fm.
We take this as an indication of numerical convergence over
the basis size in the region where the two nuclei are in contact with each other.
Before the two nuclei overlap at $R > 8$ fm,
for $N_\text{max}>80$, the results are
less stable and negative friction can appear.
This is a discretization artefact, as several
single-particle states are unbound when
$N_\text{max}>80$.

At large distances, $R>8.5$ fm, the
friction coefficients in Fig.~\ref{fig:conv}(a) are asymptotically zero.
This indicates that the two nuclei keep their ground state properties in
the approaching phase.
As the two nuclei start to overlap,
$\gamma$ first increases to $\gamma \approx 65$ c/fm at
$R \approx 5.5$ fm, and subsequently decreases
to a value of $\gamma \approx 18$ c/fm as $R \to 0$.
The hump peak at intermediate distances corresponds to the
region where collective motion is most damped.
The position of the peak turns out to depend on the
incident energy, as will be shown below.
The shape of the friction coefficient curve qualitatively
agrees with the calculations of the DD-TDHF method \cite{WLA09,WL08}.

The dependence on $R$ of the $\beta$ coefficient is very similar.
At large distances $R > 8.5$ fm, $\beta=0$,
indicating that the total energy can be conserved without
the additional term $v'$ in Eq.~(\ref{eq:tddm1}).
As the two nuclei overlap at distances below $R = 8.5$ fm,
$\beta$ starts to grow.
$\beta(R)$ presents
a maximum that coincides with the maximum of $\gamma(R)$.
This may imply that in a conventional TDDM$^P$ calculation,
a finite basis may cause inadequate dissipation.
Unlike $\gamma$, when the system gets more compact as $R$ decreases,
$\beta$ reduces to zero again. Our results thus indicate
that energy non-conserving effects are maximal in the region after
contact, when the collective motion in the compound nucleus is more strongly
damped.
We note that $\beta$ is positive throughout the evolution, which indicates that
the energy-conserving dynamics is preferentially reducing the momentum of
single-particle states.

The results in Fig.~\ref{fig:conv} validate the strategy
discussed in Sec. \ref{sec:theob}.
The convergence is achieved as the basis size increases.
The physical $\gamma$ friction coefficient is relatively
insensitive to the total basis size compared to $\beta$.
The maximum value of $\gamma$ increases
by less than $21 \, \%$ when going from $N_\text{max}=48$ to $N_\text{max}=88$.
In contrast, the adjusted parameter $\beta$ is more sensitive to the model space and
increases by a factor of $4$.
We note that $\beta \neq 0$ even with the very large basis sizes explored here,
which means the violation of energy conservation
in this implementation of TDDM can not be remedied
by increasing the number of moving orbits. 

\begin{figure}
\begin{centering}
\includegraphics[width=0.90\columnwidth]{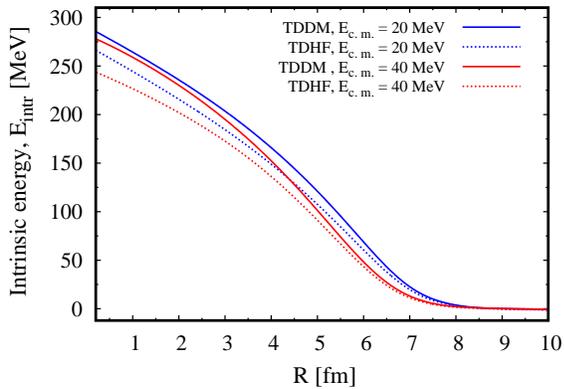}
\par\end{centering}
\caption{\label{fig:intr}(Color online)
Intrinsic energy 
as a function of $R$. 
The solid and dotted lines
indicate the results of TDDM and TDHF simulations, respectively.
The blue (upper) and red (lower) lines indicate the results at incident energies of
$E_{\rm c. m.} = 20$ MeV and $E_{\rm c. m.} = 40$ MeV, respectively.
}
\end{figure}

We now turn to look at the effect of two-body collisions in the
dissipation processes.
For comparison, we simulate the reaction at two different incident
energies $E_{\rm c. m} = 20$ MeV and $E_{\rm c. m} = 40$ MeV,
with TDDM$^P$ and TDHF, using the
same underlying mean-field interaction.
The TDDM$^P$ dynamics are computed with $N_\text{max}=60$.
Fig. \ref{fig:intr} shows the intrinsic energy defined
in Eq.~(\ref{eq:intr}) as a function of $R$ for
both TDHF and TDDM$^P$ simulations at the two incident energies.
At large distance, a zero intrinsic energy indicates again that the two nuclei
remain close to the ground state in the approaching phase.
After contact, the intrinsic energy grows monotonically
in the region $R < 8$ fm all the way to values of $E_\text{intr} \approx 250-300$ MeV
at $R\to0$.
Compared with the evolution at $E_{\rm c. m.} = 20$ MeV,
the dissipation processes at $E_{\rm c. m.} = 40$ MeV start
relatively later. A non-zero value of $E_\text{intr}$ appears
at a smaller $R$ at $E_{\rm c. m.} = 40$ MeV, corresponding to
a more compact configuration.
This feature of retarded dissipation has been studied in
TDHF \cite{WLA09}. For the TDHF simulations,
the curves at $E_{\rm c. m.} = 20$ MeV and
$E_{\rm c. m.} = 40$ MeV increase at very similar rate.
For TDDM$^P$, $E_{\rm intr}$ grows faster at higher
collision energy in the interior region $R < 4$ fm.
We take the difference between TDDM$^P$ and TDHF intrinsic energies as
an indication of dissipation stemming from two-body correlations.
The results in Fig.~\ref{fig:intr} suggest that,
the higher the incident energy is,
the more energy is dissipated from two-body
collisions. This effect can be further quantified
by looking at the friction coefficient and the $\beta$ parameter. 

 \begin{figure}
\begin{centering}
\includegraphics[width=0.90\columnwidth]{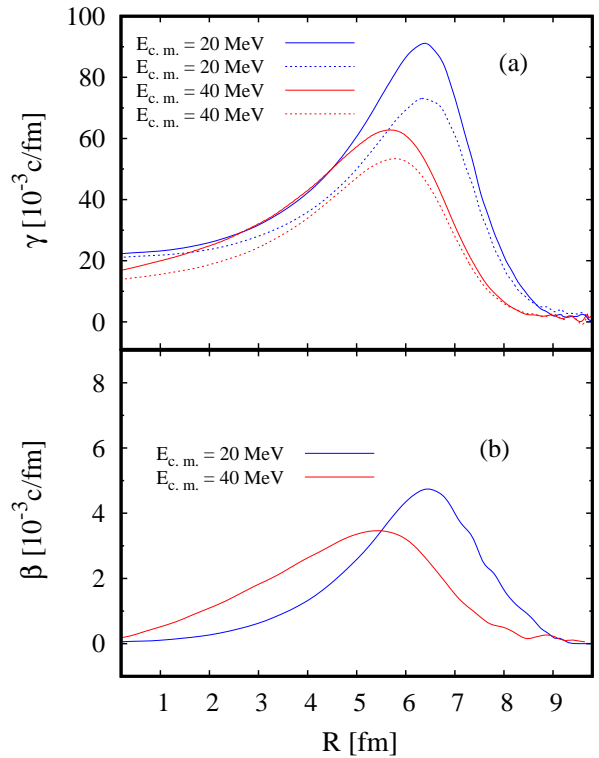}
\par\end{centering}
\caption{\label{fig:fric}(Color online)
Panel (a): friction coefficient $\gamma$ for TDDM$^P$
(solid lines) and TDHF (dotted lines) simulations as a function of $R$.
Blue (upper) and red (lower) lines indicate the results at incident energies of
$E_{\rm c. m.} = 20$ MeV and $E_{\rm c. m.} = 40$ MeV, respectively.
Panel (b): $\beta$ parameters in the same conditions.
}
\end{figure}

Figure~\ref{fig:fric}(a) shows the friction coefficient
$\gamma$ obtained in TDDM$^P$ (solid lines)
and TDHF (dotted lines) simulations.
The parameter $\beta$ in the TDDM$^P$ simulations is shown in Fig.~\ref{fig:fric}(b).
The shapes of the friction coefficients as a function of $R$ are very
similar for both the TDDM$^P$ and the TDHF simulations.
At large $R$, there is no active dissipation at either
energy, so $\gamma=0$. As $R$ decreases, the friction coefficients
develop a hump. The position and size of the maximum changes
depending on incident energy and the treatment of two-body dissipation.
The lower incident energies correspond to larger maxima at larger separations.
For instance, for TDDM$^P$ at $E_{\rm c. m.} = 20$ MeV, the peak value
is $\gamma=86$ c/fm at $R=6.5$ fm, whereas at $E_{\rm c. m.} = 40$ MeV
the corresponding maximum friction coefficient is $\gamma=64$ c/fm,
at $R=5.7$ fm. Thus, friction is more effective at lower incident energies.

It is in the peak region of $\gamma$ that the largest differences between the treatments
of correlations are found. $\gamma_{\rm TDDM}$ is about $20 \, \%$ larger than
$\gamma_{\rm TDHF}$ for the two incident energies considered here.
This shows that two-body correlations contribute to
enhance dissipation effects.
The increase in $\gamma$ due to two-body collisions is more significant
at lower incident energy.
This is at odds with the discussion around Fig.~\ref{fig:intr},
which indicated that, in terms of intrinsic excitation energy, two-body
correlations contribute more at higher energies.
The friction coefficient is however inversely proportional
to the collective momentum,
it therefore probes the time-dependence of the reaction in a more
sensitive way.
This energy dependence also agrees
with the results obtained in DD-TDHF calculations \cite{WLA09,WL08}.
We note that as $R$ becomes smaller and the compound nucleus contracts,
friction becomes less important. In the region where $R \lesssim 3$ fm, all
the results flatten out to values $\gamma \approx 20$ c/fm.

The results for $\beta(R)$ are shown in Fig.~\ref{fig:fric}(b). 
The values of $\beta$ are one order of magnitude smaller than those of $\gamma$.
$\beta$ is very close to zero both at large $R$ and small $R$.
It is positive at both incident
energies and develops a clear maximum as a function of $R$.
At lower (higher) energies, the maximum is larger (lower), with
$\beta \approx 4.8$ c/fm ($\beta \approx 3.6$ c/fm)
at $R \approx 6.5$ fm ($R \approx 5.7$ fm). The position
of the maximum of $\beta$ coincides with that of $\gamma$ at
both incident energies. We take this as an indication that
two-body dissipative effects are maximal at the point where
the incompleteness of the basis is more critical. 

Finally, we comment on the absolute values of both $\gamma$ and
$\beta$. Maximum values of $\gamma$ are about a factor of
$20$ larger than maximum values of $\beta$. Both quantities have
the same units and represent, in some way, inverse timescales associated
with dissipation. $\gamma$ can be thought of as the inverse timescale associated
with dissipative friction effects. One can interpret $\beta$ as the inverse
timescale associated with basis incompleteness (or any other mechanism) bringing
in energy non-conservation. The very small values of $\beta$ indicate that
dissipation is governed by the relatively faster friction processes, whereas any
energy non-conserving effects set in later in the dynamics. We therefore
expect that our results will hold in all implementations of TDDM, independently
of whether an incomplete moving basis is used or not.

\section{\label{sec:sum} Summary}

In this work, we study the fusion dynamics of light
nuclei using time-dependent simulations. We are interested
in quantifying the effect of dissipative effects beyond the
mean-field level. To this aim, we implement
TDHF and TDDM$^P$ simulations of the the nuclear fusion reaction
$^{16}$O+$^{16}$O $\rightarrow$ $^{32}$S.
We propose a method to remedy the problem
of energy non-conservation in the TDDM approach.
The method is based on the idea that the finite moving basis
in which the TDDM dynamics is described can be optimized
by introducing a correction term in the mean-field
Hamiltonian.
With conservation of energy restored,
we apply a macroscopic
reduction procedure to TDDM simulations to
study the dissipation mechanisms.
The friction coefficients
are extracted for both TDHF and TDDM calculations.
We find that the size of the basis does not qualitatively
affect the determination of the TDDM friction coefficients.

Compared to the results of TDHF dynamics, where
two-body correlations are absent,
we find that dissipation is enhanced
noticeably in TDDM simulations.
For instance, the friction coefficients in
TDDM dynamics can be up to $20 \, \%$ larger than
TDHF results. We also find that at higher bombarding energy,
two-body correlations provide
a larger contribution to dissipation than at lower energy.

The form of interaction is critical in TDDM
simulations, and could play an important role in the
dissipation process.
In this work, we use Skyrme III to construct the mean-field
in the TDHF and TDDM calculations.
We have used a generalized pairing and surface-dominated interaction under
the TDDM$^{P}$ approximation.
The work presented here is the first investigation in this
direction, but the choice of the interaction remains ambiguous.
A more thorough analysis of how both the mean-field and
the residual interactions affect these results would be a first
step beyond this work.
A systematic study on the dissipation process for different
reaction systems is also undergoing.


\begin{acknowledgments}

This material is based upon work supported by STFC through Grants ST/L005743/1
and ST/P005314/1.
This work is also supported (in part) by Interdisciplinary Computational Science Program in CCS,
University of Tsukuba.
\end{acknowledgments}

\bibliographystyle{apsrev4-1}
\bibliography{biblio}


\end{document}